\documentclass[12pt]{article}
\usepackage{fullpage}
\usepackage{latexsym}
\usepackage{amsmath,amsfonts,amssymb,amsthm,enumerate,dsfont,color}

\newcommand{\ignore}[1]{}

\newtheoremstyle{postnum}
  {\topsep}
  {\topsep}
  {\slshape}
  {0pt}
  {\bfseries}
  {:}
  { }
  {\thmname{#1}\thmnote{ (#3)}}

\theoremstyle{postnum}

\newtheoremstyle{prenum}
  {\topsep}
  {\topsep}
  {\slshape}
  {0pt}
  {\bfseries}
  {.}
  { }
  {\thmnumber{#2}\thmname{ #1}\thmnote{ (#3)}}

\theoremstyle{prenum}

\newif\ifnotesw\noteswtrue
   {\ifnotesw\marginpar[\hfill\(\top\)]{\(\top\)}\fi}%
   {\ifnotesw\marginpar[\hfill\(\bot\)]{\(\bot\)}\fi}

\newcommand{\mnote}[1]%
    {\ifnotesw\marginpar%
        [{\scriptsize\begin{minipage}[t]{\marginparwidth}
        \raggedleft#1%
                        \end{minipage}}]%
        {\scriptsize\begin{minipage}[t]{\marginparwidth}
        \raggedright#1%
                        \end{minipage}}%
    \fi}

\title{Roman Jackiw: A Beacon in a Golden Period of Theoretical Physics}
\author{Luc Vinet \\
\footnotesize{Centre de Recherches Math\'{e}matiques, Universit\'e de Montr\'eal, Montr\'eal, QC, Canada}\\
\tt \footnotesize{vinet@crm.umontreal.ca}}

\begin{document}
\maketitle
\begin{abstract}

This text offers reminiscences of my personal interactions with Roman Jackiw as a way
of looking back at the very fertile period in theoretical physics in the last quarter of the 20th century.
\end{abstract}

\bigskip

\textit{To Roman: a bouquet of recollections as an expression of friendship.}

\section{Introduction}

I owe much to Roman Jackiw: my postdoctoral fellowship at MIT under his supervision has shaped my scientific life and becoming friend with him and So Young Pi has been a privilege. Looking back at the last decades of the past century gives a sense without undue nostalgia, I think, that those were wonderful years for Theoretical Physics, years that have witnessed the preeminence of gauge field theories, deep interactions with modern geometry and topology, the overwhelming revival of string theory and remarkably fruitful interactions between particle and condensed matter physics as well as cosmology. Roman was a main actor in these developments and to be at his side and benefit from his guidance and insights at that time was most fortunate. Owing to his leadership and immense scholarship, also because he is a great mentor, Roman has always been surrounded by many and has thus generated a splendid network of friends and colleagues. Sometimes, with my own students, I reminisce about how it was in those days; I believe it is useful to keep a memory of the way some important ideas shaped up and were relayed. Hence as a tribute to Roman, I thought of writing the following short account of my personal connections with him in addition to the scientific hommage written in collaboration with my colleagues Nicolas Cramp\'e and Rafael Nepomechie. I fully appreciate that my own little history is of no special interest but I am offering this text as an illustrative testimony of a vibrant intellectual period in the companionship of Roman and of other scientists who like him were larger than life.

\section{First encounters with Roman and his work as a graduate student}

The Centre de Recherches Math\'ematiques broadly known as the CRM, was founded  in 1968. One only appreciates with hindsight which butterfly wing flaps will have a determining effect on your life. In my case one of these was the Prague Spring and its repression which occurred in 1968. As a result, two outstanding Tchech physicists, Jiri Patera and Pavel Winternitz educated in the highest Soviet
scientific tradition took up positions at the CRM in the following years. 
They then gave to Montreal a big research impetus and developed in this city a strong school in mathematical physics. I was fortunate to join the CRM and to pursue graduate studies within their group in the mid 70s. In theoretical physics, the end of the 60s saw the advent of the Weinberg-Salam model \cite{Wein}, \cite{Salam} unifying the electro-weak interactions, followed by the proof of its renormalizability\cite{Hooft}, the development of QCD \cite{GW, P}, which had been preceded by the discovery of the Adler-Bell-Jackiw anomaly \cite{A, BJ} (beautifully presented in \cite{J}). As a young student, I very much wished to  get involved in those striking developments in field theory. With a much enthusiastic postdoctoral fellow then at the CRM, John Harnad, who de facto became my co-supervisor, we started a gauge theory seminar. The importance of classical solutions with topological properties such as magnetic monopoles and instantons in non-perturbative analyses was being revealed at about that time. The review on monopoles by Goddard and Olive \cite{GO} as well as Coleman's Erice lectures \cite{C} were extremely formative. In those years, the Theory Division of the Canadian Association of Physicists was organizing Summer schools in Banff. This is where I met  Roman for the first time in 1977 as he delivered lectures \cite{JNR} on classical solutions of the Yang-Mills theory that much impressed me. Two papers on this topic which appeared around that time had a big influence on my Ph.D. work: the first by Roman and his collaborator Rebbi \cite{JR} (and reviewed in \cite{JNR}) where the conformal $SO(5)$ invariance of the BPST one-instanton solution is identified and the second by Witten \cite{W} where a muti-instanton solution is obtained through the dimensional reduction under $SO(3)$ of the self-dual Yang-Mills equations to a Higgs model in curved two-dimensional space. The essential aspect in these studies is that the variance of the Yang-Mills field under the space-time transformation is compensated by a gauge transformation. These publications prompted John Harnad and myself with a geometer, Steve Shnider, then at McGill University to look systematically at the description of gauge fields that are invariant in that sense under space-time transformations. We treated this problem in the global formalism of fiber bundle theory that had gained popularity amongst physicists in part thanks to a paper by Wu and Yang \cite{WY}; we classified lifts of (certain) group actions on manifolds to principal bundles over such bases and then characterized the invariant connections under the lifted actions. We then applied these results to obtain solutions to the Yang-Mills equations on compactified Minkowski space. This formed the core of my thesis and the paper on the general framework remains one of my most cited. At some point in the course of these investigations, Steve made a preliminary presentation \cite{HSV} at a conference in Lawrence, Kansas in the Summer of 1978 which focused on the twistorial approach to instantons. Using this opportunity to take his son Nick as I recall on a road trip across America, Roman was one of the principal speakers \cite{J2} at this meeting which was my first occasion to present some of my work to him. Those questions regarding invariant gauge fields were timely then and in fact were addressed simultaneously by Forgacs and Manton \cite{FM} in a complementary fashion using infinitesimal methods. As I was completing these projects, in 1979, I had the occasion to accompany my co-supervisor Pavel Winternitz during his sabbbatical to Saclay. This is how I obtained a Doctorate from the Universit\'e Pierre et Marie Curie for work I had done separately on the classification of second order differential equations in two dimensions that are invariant under subgroups of the conformal group. Upon my return to Montreal, I was wrapping up my Ph.D. and the question came as to where I should go for my postdoctoral fellowship. To explore the possibility of becoming a Research Assistant at MIT, John Harnad extended to Roman an invitation to visit Montreal and this is how a few months later I was moving to Cambridge. We especially remember from that visit that Roman was asking if we knew the three-form whose exterior derivative would give the Chern four-form. He clearly was on his way to developing the Chern-Simons gauge theories, we still regret that at the time we did not have the culture to provide the answer and join him in these investigations.

\section{The postdoctoral years at MIT}

In some sense Roman is responsible for my meeting my wife. I had become desperate to find an apartment in the Boston area and thought of checking where Roman lived. As it happened I looked in an old phone book that had 808 Memorial drive as his address. It is essentially where MIT ends and where Harvard begins. As I discovered later, Roman only stayed there briefly, but I found an apartment in that building with a nice view of the Charles. The apartment number was 512. A few months later, a charming girl named Letitia rented apartment 1212. Born in Montreal, she came to take up a research engineer position at MIT's Lincoln Lab. Meeting her has been one of the best things that have happened in my life.

When I arrived in September 1980 at MIT's Center for Theoretical Physics (CTP), Nick Manton had already been there for one year also as a postdoc and by that time had co-authored a paper \cite{JM} with Roman applying the theory of invariant gauge fields to the determination of constants of motion in their background. It was great to get to meet Nick, the disappointment came when Roman told me that he had achieved what he wanted to do in relation with my thesis work and was moving on to different problems. Having thus been taught that timing is of importance, I carried on and dove with excitement, delight and some trepidations in the amazing environment that the CTP was providing. It is not really possible to describe in a few words how vibrant the CTP was. The Faculty formed a truly outstanding group. Altogether they received numerous recognitions. During my postdoctoral fellowship alone in 1981, Jeffrey Goldstone obtained the Dannie Heineman Prize (which Roman got in 1995) and Viky Weisskopf was awarded the Wolf Prize. I also recall a meeting in honour of Francis Low who was the Provost of MIT on the occasion of his 60th birthday. This was the time when Alan Guth, a former student of Low who had joined the MIT Faculty in 1980 was developing an inflation cosmology \cite{G} and when So -Young Pi was much involved in these developments \cite{GP, Pi}. And there was Roman who was attracting like a magnet a very large number of highly talented Ph.D. students and postdocs. Many of them are actually contributing to this festschrift with Antti Niemi one of the instigators of this project. It was a privilege to be part of that group which boasted incredible creativity of intensity. Under the leadership of Roman and thanks to his inspiration many important chapters of theoretical physics were developed through various collaborations involving members of that team, among these advances are the studies of the fermion-vortex system and the fermion number fractionization, the development of the Liouville field theory, the topologically massive (Chern-Simons) gauge theories and many more. It was hard to keep track with all these papers ``typeset in TeX by Roger L. Gilson". The  `brown bag" lunch seminar at the CTP certainly created great occasions for exchanges and the generation of research ideas. The interactions with Harvard were also highly stimulating. I have much benefitted for instance of the lectures on supergravity given by Steven Weinberg on one of his extended visits. The joint Harvard-MIT theoretical physics seminar in particular was a Thursday ritual not to be missed. Much appreciated also were the dinners with the speaker that followed at which the postdocs were invited. 
The meals at the Yenching chinese restaurant near Harvard square with Sidney Coleman placing the order bring fond memories. I keep saying that this is where I learned to use chopsticks, out of necessity,  since I did not manage to eat much because of my poor dexterity the first few times I attended. In 1981, my second year at the CTP, Eric D'Hoker arrived at MIT for a three-year postdoctoral fellowship and so did Eddie Farhi to take up a Junior Faculty appointment. Eddie and Roman edited a book at that time on dynamical gauge symmetry breaking \cite{FJ}.  Eric and I became close friends and began collaborating on problems that were often prompted by Roman. Thanks to the `Bruno Rossi" exchange program between the INFN and MIT and Sergio Fubini in particular, the CTP entertained close ties with Italian physicists. In a paper published in 1976, de Alfaro, Fubini and Furlan have explored the ramifications of scale invariance in one-dimensional quantum mechanics \cite{DFF}; building on this work in a more physically motivated context, Roman identified in 1980 the conformal symmetries of the magnetic monopole \cite{J3}. At some point Roman passed on to Eric and I manuscript notes from Fubini that led to the paper with Rabinovici developing superconformal quantum mechanics \cite{FR} and which stemmed from a lot of on-going interest in supersymmetric theories (see for example \cite{W1}). That led us to consider the supersymmetries of monopole systems which we did (see for instance\cite{DV1}) by extending the work of Roman. We carried on studies \cite{DV2} that built on the seminal work of Deser, Jackiw and Templetion on topologically massive theories \cite{DJT}; we did work then on reduction of higher dimensional Chern-Simons theories that for some reason we never published. 

Those postdoctoral years at MIT largely thanks to Roman were indeed defining ones. When I left Cambridge and the end of my fellowship, I promised myself to return as soon as possible.

\section{Returning to MIT as a young Faculty at Universit\'e de Montr\'eal} 

In 1982, holding University Research Fellowship from NSERC, I came back to my hometown to take on an Assistant Professor position at the Universit\'e de Montr\'eal. I kept reaping the benefits of my sojourn at MIT: the first postdoc I hired was Haris Panagopoulos  whom I had met there; I arranged for Manu Paranjape a fellow Canadian and former student of Goldstone with whom I also had become acquainted at the CTP to come to Montreal as Assistant Professor at UdeM and I kept working with Eric D'Hoker as he moved himself to Columbia University.  

In 1987 as I was presenting my dossier for tenure, I held the promise I had made to myself and returned to the CTP as Visiting Researcher for six months at the beginning of the year. I then shared an office with Oscar Eboli, a most charming Brazilian fellow who was working with Roman and So-Young on quantum fields out of thermal equilibrium \cite{EJP} and with whom I had the pleasure to reconnect in Sao Paulo in the Summer of 2018.

At the time So-Young and Roman's son Stefan was two years old.

Shortly after my arrival, Roman introduced me to a brilliant young Italian physicist named Roberto Floreanini. The two were working on the functional approach to quantum field theories \cite{FHJ}. It clicked immediately with Roberto. I embarked in that program and co-authored a few articles (e.g. \cite{FV1})  with Roberto before I returned to Montreal. Once again, because of MIT, I had not only found a superb collaborator but met someone who became a dear friend. And so Roberto and I collaborated intensively until to his disappointment I suppose, I became Provost at McGill University.

As always, it was truly enriching scientifically and personally to be around Roman. I was living during that stay at Longfellow Place near So-Young and Roman's home in Beacon Hill. I recall one evening when Roman was giving me a lift in his BMW with the iconic ``FFDUAL" plates, the radio was on and he was quite taken by the music saying how beautiful it was. He asked : do you know what it is? I was not completely sure but I thought I had recognized Richard Strauss and said so. A few minutes after I had passed my door, I have a phone call, it was Roman to tell me: you were right it was Ariadne auf Naxos. A cute anecdote to recall how classy Roman is.

At the time in the first half of 1987, Roman was also examining issues connected to Berry potentials \cite{J4}. He shared with me the thought that the symmetries of a problem could determine the associated Berry connections. By then, I was back in Montreal. It was quite nice to think that the theory of invariant connections of my PhD days could be brought to bear on this current topic. I sorted this out and wrote a draft.
At this point I had not co-authored papers with Roman. I then thought the moment had come to loop the loop and publish something with Roman on the topic that had initially brought me to MIT. In spite of my insistence, Roman declined arguing that I should publish the paper alone \cite{V}. He subsequently wrote an article \cite{J5} covering the question in his own way and kindly quoted my publication. Although this left me with the regret of never having had the pleasure of chiseling a text with Roman this turned out to be some kind of blessing for me and (I apologize Roman) a certain curse for him. Indeed as you are all well aware, in any grant application process one needs to suggest prestigious colleagues with whom you have never collaborated (or at least not within a certain period). I have abused of Roman in this respect but magnanimously he always obliged and has been very supportive. 

During the academic year 1989-1990, Eric D'Hoker kindly hosted me at UCLA where he had moved. Our first son, Jean-Fran\c{c}ois, was born in LA in July 1990. Roberto Floreanini visited and this is when we launched as we were babysitting what became a vast study of the connections between quantum algebras, q-special functions and their applications.

\section{Meeting here and there}

The occasion for another prolonged visit at the CTP never presented itself again. Nevertheless life, family and common scientific interests provided opportunities for So-Young, Roman, Letitia and I to get together sometimes with our children and those were always  very happy moments. Our friendship built over the years but I think that it certainly strengthened while we were all together in Banff in 1989.  I have mentioned before that these Summer schools in the Canadian Rockies which provided my first meeting with Roman as a student, were an institution. This one entitled Physics, Geometry and Topology had a stellar group of lecturers including Roman who spoke about planar physics \cite{J6}. Although very successful it turned out to be the last school of that nice series because the original sources of funding disappeared. I had enjoyed them so much that I had the idea to revive them under the name \textit{CRM Summer School in Banff} when I became director of the CRM in the nineties. This turned out to be the precursor of what is today the highly successful Banff International Research Station (BIRS) which is jointly supported by the CONACYT, the NSF and NSERC.

A peculiar encounter with Roman and one that brings laughs in retrospect occurred in Kiev in 1992. There was a conference organized by the Ukrainian Academy of Sciences. Roman was participating to
connect with his childhood days I believe, and I was attending  because this conference was taking place immediately after a meeting in Alushta, Crimea at which I had been invited by Alexei Morozov and other friends from ITEP in Moscow. Phong who is a member of the Department of Mathematics at Columbia University and a collaborator of Eric D'Hoker was also in Crimea. Based on our understanding of what ikra meant in Russian, Phong and I were proudly thinking that we had managed to buy nice caviar. And so I arrived in Kiev with two tin cans of this great finding. My father who had wished to visit Crimea was accompanying me. Once in our room in the hotel of the Academy we noticed that there was no hot water. We got together with Roman at some point to confirm that he did not have hot water either because this time in the Summer had been chosen by the hotel management to perform plumbing work throughout the building. Facing this adversity, we told Roman about our caviar and invited him in our room with the hope of indulging in this delicacy only to find after some struggle to open the cans that they contained some dry and highly salted red fish eggs. So much for our refined party!

We have also met in less exotic places like Boston or New York and at times our reunions were prompted by musical reasons. Stefan Jackiw is a magnificent violonist who is unanimously recognized as belonging to an elite group of only a few. My son Jean-Fran\c{c}ois, has done musical performance training as a violist to a high level even though he chose not to pursue a professional career.
The 2006 edition of the International Conference on Group Theoretical Methods in Physics (ICGTMP) was held in New York. I took along Jean-Fran\c{c}ois and his younger brother Laurent under the condition that they visit one museum every day, an assignment that they fulfilled. Stefan was already living in New York and So-Young and Roman who were also staying in Manhattan at that moment very kindly arranged for the six of us to have a joyful dinner after Stefan had kindly practiced scales with Jean-Fran\c{c}ois. In 2009, So-Young organized a gathering for Roman's 70th birthday in Boston which was a lovely event and another occasion to celebrate Roman's outstanding impact on science and people.  For five years in a row, beginning in 2010, Jean-Fran\c{c}ois participated in the Aspen Music Festival and School. We have always managed to spend some family time in Aspen during those Summers. 
In some of those years Stefan has been a guest artist. I recall in particular being subjugated one evening by an interpretation he gave of the third Brahms sonata. So-Young and Roman were also involved regularly in the Aspen Center for Physics and our stays often overlapped. We never missed the chance to get together and I vividly remember great dinners at the Pine Creek Cookhouse at the base of the Elk Mountains.

\section{Roman and the CRM and Montreal}

As I bring these reminiscences to a close, I want to stress how generous Roman has been with his visits to Montreal and the connections he has built with this city and its scientific organizations. I shall point at three moments in particular.

In 1988, Yvan Saint-Aubin and I organized the XVIIth ICGTMP in Montreal. This international conference took place against the backdrop of perestroika which created  very fortunate circumstances that                                                                              
allowed proeminent Soviet mathematicians and scientists such as Belavin, Faddeev, Fatteev, Manin and Zamolodchikov to attend and lecture at the meeting. For many, this was their first visit to North
America. The list of plenary speakers was stunning and we were lucky to have Roman \cite{J7} (who also gave an additional talk \cite{J8}) and So-Young \cite{Pi2} among them. The recipient of the 
Wigner medal awarded during the conference was Isadore Singer from MIT whose celebrated index theorems are so intimately connected with Roman's work. This made for a superb program; the 
conference was very successful and many of the MIT friends (D'Hoker, Eboli, Floreanini, Niemi, Panagopoulos, etc) attended. 

For a while Roman enjoyed smoking little cigars and I wondered if this was not what explained why he appreciated Montreal: cuban cigars could be found in this trendy Canadian city! Indeed whenever in
town Roman would make sure to stop at the Davidoff store to stock up. There was thus a time when as a friendly gesture, I would make sure of smuggling some havanas on trips to Boston. For the good
of Roman's health and my standing with the US customs officers there is now prescription on these infractions.

In 1993, I was appointed Director of the CRM (for the first time). As a central part of its activities are thematic programs. These concentration periods on topics of special interest bring specialists from all 
over the world around a number of workshops and conferences; these are planned with significant leadtime. The Aisenstadt Chair is CRM's most prestigious lectureship: it is offered to distinguished 
scientists upon the recommendation of the CRM International Advisory Committee. The holders of the Chair deliver series of talks that are integrated within the thematic program
of the semester or of the year; they are also strongly encouraged to turn their lectures into a book to be published in one of the CRM monograph series. At the beginning of my second term as Director 
around 1997, together with Philippe Di Francesco, Lisa Jeffrey, Andr\'e Leclair and Yvan Saint-Aubin we started putting together a theme year in mathematical physics. Little did I know that I would be
appointed Provost of McGill University at the beginning of July 1999. Even though I could not enjoy as much as I had intended the deployment of the scientific events that the year entailed, the program was a
resounding success with Roman holding the Aisenstadt Chair. He gave his lectures in the framework of two workshops: the first on Strings, Duality and Geometry and the second in Condensed Matter and
 Non-Equilibrium Physics. The general topic he chose was Fluid mechanics and as you may imagine Roman offered an original and fascinating view of this broad subject from the perspective of a particle
 theorist \cite{J9}.
 
 Being Provost at McGill led to my becoming Rector (or President) of the Universit\'e de Montr\'eal and so I went around the Mont Royal returning to the institution where I had begun my academic career.
 Time was at a premium in those years but as mentioned above the Jackiws and the Vinets kept meeting and I managed to maintain some research activity. I was determined not to end my term as Rector 
 before ensuring that Roman receives recognition from us for his immense scientific accomplishments and his special relation with Montreal. This happened in 2010 when I had the great pleasure to
 present him with a Honorary Doctorate from the Universit\'e de Montr\'eal. This was a touching celebration that took place within the solemn Ph.D. convocation in a packed amphitheater. Roman with his
 usual intellectual elegance gave an inspiring acceptance speech and generated a long and enthusiatic applause. I recall that as he was leaving the podium looking quite moved he said to me: Now I know
 how my son feels!

\section{Envoi}

Dear Roman:

I see you like a great artist. Writing this short and sketchy chronicle of our encounters over the years gave me the chance to reflect further on your work and its tremendous impact. Your papers 
and expository texts are like magnificent paintings that reveal subtle and unexpected perspectives. These paintings were much acclaimed when they were first presented and brought you fame. Apprentices came to learn from you and many emulated you from the distance. You generously shared your knowledge and craftmanship. If you had such an influence on me, we can imagine the number of people for whom this has been the case. And then new generations rediscover your work, look at it from different angles, apply it to different phenomena, other artists get inspired by it and create new movements, avatars of your past creations. You engage with that and produce new work making the wheel turn. 

Need it be said that we look forward to more of these inspiring pieces and to your advice and views moving ahead. Thank you for the past and thank you for the future.
Here is to you from the mind and from the heart.

{\paragraph{Acknowledgements} 
I wish to acknowledge that much of my connections with Roman for which I am so grateful and of which I have given only a poor glimpse here, have been made possible by the Natural Science and Engineering
Research Council (NSERC) of Canada. Over the years NSERC has offered me personally postgraduate scholarships, a postdoctoral fellowship and various discovery grants for which Roman relentlessly wrote recommendations. NSERC has also continuously supported this marvellous institute for research in the mathematical sciences that the Centre de Recherches Math\'ematiques or CRM is and that I still have the privilege to lead; this has in particular allowed our community to interact so profitably with Roman Jackiw at various occasions. To NSERC and through this Council to the Canadian taxpayers who are making scientific journeys possible: thank you.}

\end{document}